\documentclass[letterpaper,twocolumn,10pt]{article}
\usepackage{usenix}

\usepackage{subfigure}
\usepackage{tikz}
\usepackage{amsmath}

\usepackage{filecontents}

\usepackage{booktabs} 
\usepackage{algorithm}
\usepackage{algpseudocode}
\usepackage{listings}
\graphicspath{{figure/}{figures/}}
\usepackage{tcolorbox}
\tcbset{sharp corners, boxsep=0pt, left=5pt, right=5pt, before skip=-1mm, after skip=-1mm}

\definecolor{edgeblue}{RGB}{0, 0, 200}
\definecolor{edgegreen}{RGB}{0, 200, 0}
\definecolor{gptgreen}{RGB}{0, 166, 126}
\definecolor{scholarpurple}{RGB}{169, 1, 251}
\definecolor{bgcode}{rgb}{0.95,0.95,0.95}

\usepackage{xcolor}

\definecolor{kscolor}{rgb}{0.9,0.1,0.1}
\definecolor{mscolor}{rgb}{0.1,0.1,0.9}
\definecolor{stcolor}{rgb}{0.1,0.9,0.1}

\usepackage{listings, xcolor}

\definecolor{verylightgray}{rgb}{.97,.97,.97}

\lstdefinelanguage{Solidity}{
	keywords=[1]{anonymous, assembly, assert, balance, break, call, callcode, case, catch, class, constant, continue, constructor, contract, debugger, default, delegatecall, delete, do, else, emit, event, experimental, export, external, false, finally, for, function, gas, if, implements, import, in, indexed, instanceof, interface, internal, is, length, library, log0, log1, log2, log3, log4, memory, modifier, new, payable, pragma, private, protected, public, pure, push, require, return, returns, revert, selfdestruct, send, solidity, storage, struct, suicide, super, switch, then, this, throw, transfer, true, try, typeof, using, value, view, while, with, addmod, ecrecover, keccak256, mulmod, ripemd160, sha256, sha3}, 
	keywordstyle=[1]\color{blue}\bfseries,
	keywords=[2]{address, bool, byte, bytes, bytes1, bytes2, bytes3, bytes4, bytes5, bytes6, bytes7, bytes8, bytes9, bytes10, bytes11, bytes12, bytes13, bytes14, bytes15, bytes16, bytes17, bytes18, bytes19, bytes20, bytes21, bytes22, bytes23, bytes24, bytes25, bytes26, bytes27, bytes28, bytes29, bytes30, bytes31, bytes32, enum, int, int8, int16, int24, int32, int40, int48, int56, int64, int72, int80, int88, int96, int104, int112, int120, int128, int136, int144, int152, int160, int168, int176, int184, int192, int200, int208, int216, int224, int232, int240, int248, int256, mapping, string, uint, uint8, uint16, uint24, uint32, uint40, uint48, uint56, uint64, uint72, uint80, uint88, uint96, uint104, uint112, uint120, uint128, uint136, uint144, uint152, uint160, uint168, uint176, uint184, uint192, uint200, uint208, uint216, uint224, uint232, uint240, uint248, uint256, var, void, ether, finney, szabo, wei, days, hours, minutes, seconds, weeks, years},	
	keywordstyle=[2]\color{teal}\bfseries,
	keywords=[3]{block, blockhash, coinbase, difficulty, gaslimit, number, timestamp, msg, data, gas, sender, sig, value, now, tx, gasprice, origin},	
	keywordstyle=[3]\color{violet}\bfseries,
	identifierstyle=\color{black},
	sensitive=true,
	comment=[l]{//},
	morecomment=[s]{/*}{*/},
	commentstyle=\color{gray}\ttfamily,
	stringstyle=\color{red}\ttfamily,
	morestring=[b]',
	morestring=[b]"
}

\definecolor{codegreen}{rgb}{0,0.6,0}
\definecolor{codegray}{rgb}{0.5,0.5,0.5}
\definecolor{codepurple}{rgb}{0.58,0,0.82}
\definecolor{backcolour}{rgb}{0.95,0.95,0.92}

\lstdefinestyle{mystyle}{
    backgroundcolor=\color{backcolour},   
    commentstyle=\color{codegreen},
    keywordstyle=\color{magenta},
    numberstyle=\tiny\color{codegray},
    stringstyle=\color{codepurple},
    basicstyle=\ttfamily\footnotesize,
    breakatwhitespace=false,         
    breaklines=true,                 
    captionpos=b,                    
    keepspaces=true,                 
    numbers=left,                    
    numbersep=5pt,                  
    showspaces=false,                
    showstringspaces=false,
    showtabs=false,                  
    tabsize=2
}

\newcommand{\tool}{\textsc{Llm4Fuzz}}
\definecolor{clcolor}{rgb}{0.5,0.7,0.9}
\definecolor{kscolor}{rgb}{0.9,0.1,0.1}
\definecolor{rbcolor}{rgb}{0.7,0.4,0.7}
\definecolor{nkcolor}{rgb}{0.4,0.7,0.7}
\definecolor{rfcolor}{rgb}{0.56, 0.0, 1.0}
\definecolor{cscolor}{rgb}{0.1, 0.4, 1.0}
\definecolor{jpcolor}{rgb}{0.36, 0.54, 0.66}

\begin{document}

\date{}

\title{\Large \bf \textsc{Llm4Fuzz}: Guided Fuzzing of Smart Contracts\\ with Large Language Models}

\author{
{\rm Chaofan\ Shou}\\
UC Berkeley
\and
{\rm Jing Liu}\\
UC Irvine
\and
{\rm Doudou Lu}\\
Fuzzland Inc.
\and
{\rm Koushik Sen}\\
UC Berkeley
} 

\maketitle


\begin{abstract}
As blockchain platforms grow exponentially, millions of lines of smart contract code are being deployed to manage extensive digital assets. However, vulnerabilities in this mission-critical code have led to significant exploitations and asset losses. Thorough automated security analysis of smart contracts is thus imperative. This paper introduces \tool\ to optimize automated smart contract security analysis by leveraging large language models (LLMs) to intelligently guide and prioritize fuzzing campaigns. While traditional fuzzing suffers from low efficiency in exploring the vast state space, \tool\ employs LLMs to direct fuzzers towards high-value code regions and input sequences more likely to trigger vulnerabilities. Additionally, \tool\ can leverage LLMs to guide fuzzers based on user-defined invariants, reducing blind exploration overhead. Evaluations of \tool\ on real-world DeFi projects show substantial gains in efficiency, coverage, and vulnerability detection compared to baseline fuzzing. \tool\ also uncovered five critical vulnerabilities that can lead to a loss of more than $\$247$k.
\end{abstract}

%
%


\maketitle

\section{Introduction}

As decentralized applications and blockchain platforms experience exponential growth\cite{defigrowth}\cite{blockchaingrowth}, millions of lines of smart contract code to manage billions of dollars in digital assets\cite{tvl1}\cite{tvl2} are deployed. Regrettably, numerous high-profile attacks have exploited vulnerabilities in this mission-critical code, resulting in significant losses through logic bugs that auditors overlooked. Therefore, it is imperative to perform exhaustive security analyses of smart contracts before deployment.


Traditional manual auditing of vast smart contract codebases is error-prone and often overlooks corner-case flaws. The industry increasingly uses automated methods like testing, dynamic analysis, and formal verification to overcome these limitations. Guided fuzz testing\cite{survey1}\cite{survey2}\cite{survey3} stands as one of the most prevalent and reliable techniques employed by smart contract developers and auditors. Fuzzing is a heuristic search algorithm for finding test inputs (or test cases) that reveal vulnerabilities. A fuzzer keeps a collection of interesting test cases and mutates them individually to produce new test cases, subsequently running the software on those test cases. By generating and executing these test cases thousands of times per second, fuzzers are adept at covering code regions and discovering complex vulnerabilities. 


Still, fuzzing for smart contracts has limitations because fuzzers do not understand the semantics of the target. Due to this, much time is wasted on exploring code regions (i.e., mutating test cases that cover the code regions) that are well-tested and unlikely to contain vulnerabilities. Additionally, complex if-statements make certain code regions much harder to cover fully. Yet, fuzzers allocate the same amount of effort to exploring easy and hard-to-cover code regions, reducing their efficacy. Lastly, smart contracts are stateful software. To trigger a vulnerability, one needs a sequence of inputs (consisting of a function call and the arguments) instead of a single input. Each input potentially mutates the state, and subsequent input depends on the mutated state. Creating such a sequence is non-trivial for fuzzers as they need to ensure every input in the sequence is valid, and the order of the sequence is correct\cite{ityfuzz}.

Recent developments in LLMs have shown their ability to find vulnerabilities in smart contracts and develop potential exploits\cite{llmok}\cite{llmok2}\cite{gptscan}\cite{smartest}. However, directly asking LLMs about the existence of vulnerabilities commonly suffers from high false positives and negatives\cite{llmfails}, restricting their broader adoptions. Yet the accumulated knowledge and experience encoded within LLMs can be harnessed for a human-like code semantics analysis. This paper introduces a novel methodology, called \tool, to optimize smart contract fuzzing with LLMs. \tool\ uses LLM-generated metrics to orchestrate and prioritize the exploration of certain code regions of the target in fuzzing campaigns. 

Specifically, we demonstrate that the LLMs are capable of producing metrics like complexity and vulnerability likelihood of code regions quite accurately. We also show that LLMs can analyze \emph{invariants} (manually inserted assertions) in the code and identify invariant-related code regions.  The fuzzers can utilize these metrics to allocate more effort to exploring more fruitful code regions. In addition, we illustrate that LLMs can also identify potentially interesting sequences of function calls. By prioritizing exploring these sequences, fuzzers can efficiently find and reach interesting and vulnerability-leading states. 


To harness the potential of LLMs, we extract a hierarchical representation of the smart contract, including source code, control flow graphs, data dependencies, and metrics produced by static analysis. These elements allow LLMs to perform semantics analysis, compare historical vulnerability patterns, and pinpoint potentially interesting transaction sequences. We embed this representation and our specific goals as prompts, allowing the LLMs to generate metrics for basic blocks and functions. This information is then encoded into schedulers of fuzzers, guiding them to explore test cases based on this refined prioritization. 

We evaluated \tool\ on real-world complex smart contract projects with and without known vulnerabilities. \tool\ gains significantly higher test coverage on these contracts and can find more vulnerabilities in less time than the previous state-of-the-art. While scanning 600 live smart contracts deployed on the chain, \tool\ identified critical vulnerabilities in five smart contract projects, which can lead to significant financial loss.   




\section{Background}
\subsection{Smart Contracts}

Smart contracts are programs deployed on blockchain networks that execute autonomous digital agreements. Acting as self-executing contracts, they encode complex business logic and handle extensive financial assets without centralized intermediaries. For instance, a smart contract might facilitate, verify, or enforce a negotiation or performance of a contract, such as an automatic payment upon receipt of goods.

These contracts accept transactions from distributed participants to alter the contract state on the immutable ledger. Popular platforms for smart contract development include Ethereum, Solana, and Cardano. In this paper, we focus on Ethereum smart contracts only. Developers often use languages like Solidity and Vyper, which compile bytecode to be executed on Ethereum virtual machines.

Take the example of a decentralized finance (DeFi) application that enables users to lend and borrow cryptocurrency. A smart contract can be created to hold collateral and release funds when certain conditions are met, such as a borrower repaying the loan. This operation is governed by code, which automatically enforces the agreement without a traditional financial institution overseeing the transaction.

The persistent and decentralized nature of smart contracts necessitates thorough security analysis before deployment, given that any vulnerability can lead to a significant loss of funds. For instance, in the DAO attack\cite{attacksurvey} on the Ethereum network in 2016, a flaw in a smart contract led to the unauthorized withdrawal of over \$60 million. The code's vulnerability lay in a recursive calling bug, exploited by an attacker to drain the funds.

Below, we introduce a set of commonly used services available on Ethereum or Ethereum-compatible chains implemented as smart contracts. These services are widely used inside attacks. Knowledge of these services is crucial to understanding the motivating examples and vulnerabilities we have found. 

\textbf{Tokens} One prominent feature of Ethereum is its support for tokens, which can be thought of as digital assets or units of value built on top of the Ethereum blockchain using a smart contract. These tokens can represent a wide array of digital goods, including, but not limited to, virtual currency, assets, voting rights, or even access rights to a particular application. The ERC-20 token standard~\cite{erc20}, prevalent on the Ethereum blockchain, defines a consistent set of rules for fungible tokens. Key methods like \verb|balanceOf(address _owner)| (querying account balance of token), \verb|transfer(address _to, uint256 _value)| (transferring the token to another account), and others ensure uniform interactions across token contracts.

\textbf{Liquidity Pools} facilitate token exchanges or "swaps" in a trustless manner. A common implementation of a liquidity pool is Uniswap V2 Pool~\cite{uniswapv2}, which is a contract holding two tokens, and the price of each token in terms of the other is determined by the ratio of the quantity of these tokens in the pool. This automated market maker (AMM) mechanism follows the formula $x \times y=k$, where $x$ and $y$ are the quantity of the two tokens in the pool, and $k$ is a constant value. Users who wish to swap one token for another interact with the respective pool. The constant $k$ ensures that as the amount of one token in the pool increases, the amount of the other decreases, maintaining a balance and adjusting prices accordingly. In liquidity pools, liquidity providers like project owners supply two tokens to the pool to enable others to trade against this pooled liquidity. In return for their contribution, LPs often receive rewards.


\subsection{Feedback-driven Fuzzing}

\begin{algorithm}
\caption{Feedback-Driven Mutation Fuzzing}
\begin{algorithmic}[1]
\Procedure{FeedbackDrivenFuzzing}{$program$}
    \State $corpus \gets \text{initialize with seed inputs}$
    \State $covMap \gets \text{initialize empty coverage map}$
    \While{within time budget}
        \State $t \gets \text{select from } corpus$
        \For{$0..Energy(t)$}
        \State $mutant \gets \text{mutate } t$
        \State $newCoverage \gets program(mutant)$
        \If{$newCoverage \text{ is novel or interesting}$}
            \State $corpus \gets corpus \cup \{mutant\}$
            \State $covMap \gets \text{update with } newCoverage$
        \EndIf
        \EndFor
    \EndWhile
\EndProcedure
\end{algorithmic}
\label{alg:feedbackfuzzing}

\end{algorithm}

Fuzzing is an automated testing technique where inputs are generated randomly. Feedback-driven fuzzing\cite{aflpp} incorporates real-time feedback from the program execution on the randomly generated inputs. Instead of blindly generating test cases, feedback-driven fuzzing monitors the execution of the program. By tracking metrics such as code coverage, branch execution, or particular conditions within the code, it intelligently guides the generation of subsequent test cases that try to maximize the metrics.  The actual generation of new test cases is performed by mutating an existing test case.  This dynamic adaptation makes the fuzzing process more targeted and efficient.

Feedback-driven fuzzing typically uses power scheduling\cite{powerscheduling}\cite{powerschedule} to allocate more time to exploring favored mutants of test cases. Energies are calculated for each test case, which quantifies the favoriteness of each test case. To calculate the energy, fuzzers commonly rely on the execution time of the test case, test case length, and how rare (determined by the hit rate) the path executed by the test case is. 

The pseudocode of the feedback-driven fuzzing algorithm with power scheduling is shown in Algorithm~\ref{alg:feedbackfuzzing}. It begins by initializing a corpus and coverage map (Lines 2-3). Within a loop (Lines 4-14), it selects a test case (Line 5) and queries its energy (Line 6). Then, for an "energy" amount of time, the algorithm mutates the test case (Line 7), executes the mutated test case (Line 8), and updates the corpus and coverage map if new or interesting paths are found (Lines 9-11). The loop continues until the time budget of the campaign has been exceeded. In general, this algorithm can efficiently guide the fuzzing to uncover code vulnerabilities by prioritizing exploring interesting test cases. 

The algorithm can be used for both fuzzing traditional software and smart contracts. For smart contracts, a test case would be a sequence of inputs (i.e., transaction calls) and mutation could be inserting new input into the sequence, removing inputs from the sequence, or mutating individual input. However, traditional feedback-driven fuzzing using test coverage is ineffective for testing smart contracts due to the stateful nature of smart contracts. Test coverage is not a good metric showing what input shall be followed by another input. Existing smart contract fuzzers like \textsc{Smartian}~\cite{smartian} leverage dataflow analysis to determine the "sequential interestingness" (how interesting a sequence is). It is deemed interesting when a sequence leads to a unique dataflow pattern. \textsc{ItyFuzz}~\cite{ityfuzz} similarly introduces custom dataflow and comparison waypoints\cite{fuzzfactory} (runtime metrics providers) to determine state interestingness. More interesting states (derived from a sequence of inputs) are prioritized for exploration.  




\subsection{Large Language Models}
Derived from the Transformer architecture~\cite{llmsurvey}, models like GPT-4\cite{gpt4} are characterized by their vast number of parameters (often in the billions or trillions), enabling them to capture intricate patterns in language. Trained using unsupervised learning on extensive textual datasets, LLMs refine their parameters by predicting subsequent words in sequences, thereby gaining proficiency in language structure, semantics, and context. Recent works have applied LLMs to boost fuzzing traditional software
\cite{llm1}\cite{llm2}\cite{llm3}\cite{llm4}\cite{llm5}\cite{llm6} serving as test case generators and mutators. However, their potential to integrate deeply into the fuzzer or enhance smart contract security analysis remains unexplored. 

\section{Motivating Example}

\subsection{The AES Exploit}

The following code segment is taken from a well-known smart contract project AES, which has been exploited, and \$62k worth of assets have been stolen \cite{aes}. 

\begin{lstlisting}
function sellTokenAndFees(
    address from, 
    address to, 
    uint256 amount
) internal {
    uint256 burnAmount = amount.mul(3).div(100); 
    uint256 otherAmount = amount.mul(1).div(100); 
    amount = amount.sub(burnAmount);
    swapFeeTotal = swapFeeTotal.add(otherAmount);
    super._burn(from, burnAmount);
    super._transfer(from, to, amount);
}

function distributeFee() public {
    ...
    super._transfer(uniswapV2Pair, technologyWallet, swapFeeTotal);
    ...
    swapFeeTotal = 0;
}
\end{lstlisting}

The function \verb|sellTokenAndFees| is called every time users attempt to sell this token (checked by whether the token is sent to the liquidity pool). This function takes a fee (deducted from the amount the recipient receives) 
and adds the fee value to the state variable \verb|swapFeeTotal|, which records the profit of the project owner. Intuitively, project owners take tax from selling their tokens. \verb|distributeFee| can be called by everyone to distribute the \verb|swapFeeTotal| amount of token from the liquidity pool to project owners. This could lead to a price inflation attack. An attacker can transfer some token to the liquidity pool (recognized as selling tokens by the above code even though it is not) and ask the liquidity pool to refund (supported by the liquidity pool to prevent accidental transfer). Suppose the attacker initially has $v$ amount of token. After this operation, they are left with $v - \verb|burnAmount|$ token, as the selling fee is deducted. 
Then, the attacker can call \verb|distributeFee| to transfer out a significant amount of token ($2\times \verb|burnAmount|$) from the liquidity pool, making the pool imbalanced, which could result in price inflation of the token. 

To successfully inflate the token price such that the earnings could be greater than the cost of inflating it, an attacker needs to borrow the token and use at least eight inputs with correct order in a sequence. Each input requires valid arguments that the fuzzer takes time to find. To generate such a profitable exploit, a na\"ive stateful fuzzer would make on average $22^{8}C$ attempts, where $C$ is the average amount of attempts to get valid arguments for one input, as there are 22 candidate functions to call. Even with more advanced fuzzers like \textsc{ItyFuzz} and \textsc{Smartian} with complex heuristics, it still could not be uncovered within a reasonable time. 


\begin{lstlisting}
uint private unlocked = 1;
modifier lock() {
    require(unlocked == 1, 'Pancake: LOCKED');
    unlocked = 0;
    _;
    unlocked = 1;
}

function mint(address to) external lock returns (uint liquidity) {
    ...
}

function burn(address to) external lock returns (uint amount0, uint amount1) {
    ...
}

function swap(uint amount0Out, uint amount1Out, address to, bytes calldata data) external lock {
    ...
}
\end{lstlisting}

Another interesting observation is that \textsc{ItyFuzz}\cite{ityfuzz} and \textsc{Smartian}\cite{smartian} are significantly more interested in the liquidity pools of the project. After inspecting the power scheduling process, we identify the reason to be that the Uniswap V2 liquidity pool contract uses a state variable ``unlocked'' (Line 1) to prevent reentrancy attacks, which disallows external calls made by such a contract to call certain functions in it again. Specifically, the contract reads the ``unlocked'' at the start of the execution. If it is false, then the execution will fail (Line 2). Then, it sets ``unlocked'' as false at the start and true at the end (Line 3-6). \textsc{ItyFuzz} considers there are dataflows among functions using ``unlocked''. Thus, it would prioritize analyzing those functions and their sequences even though such a data flow is meaningless and exists between almost every function in the liquidity pools.   

\subsection{Challenges}

To summarize, the traditional fuzzing approach faces challenges working on projects existing following characteristics:

\begin{enumerate}
    \item \textbf{Distractions} The project under test may be a system of tens or even hundreds of smart contracts with thousands of public functions. Traditional fuzzers cannot identify and allocate more energy to the most vulnerable functions.

    \item \textbf{Traps} Traditional fuzzers also waste effort on functions that might be well-tested and have no vulnerabilities. One example is projects with Uniswap liquidity pools. The contracts in Uniswap liquidity pools are well-fuzzed and manually audited. Therefore, there is no need to allocate the same amount of energy to those contracts as project-specific contracts that are poorly tested. Traditional fuzzers may also get trapped in functions that are not complex but impossible to cover (e.g., functions that can be only invoked by certain accounts). Those functions are assigned the same or even more energy than functions explorable by fuzzers.


    \item \textbf{Complex Sequential Dependencies} To trigger most vulnerabilities or find exploits today, the fuzzer has to be able to produce long, interesting input sequences. Traditional fuzzers leverage dataflow information or comparison information to determine the sequences. However, in real-world projects, using solely that information can generate millions of "interesting" sequences that are not actually interesting semantically. 
\end{enumerate}

\section{Methodology}

\begin{figure*}[htbp]
  \centering
  \includegraphics[scale=0.51]{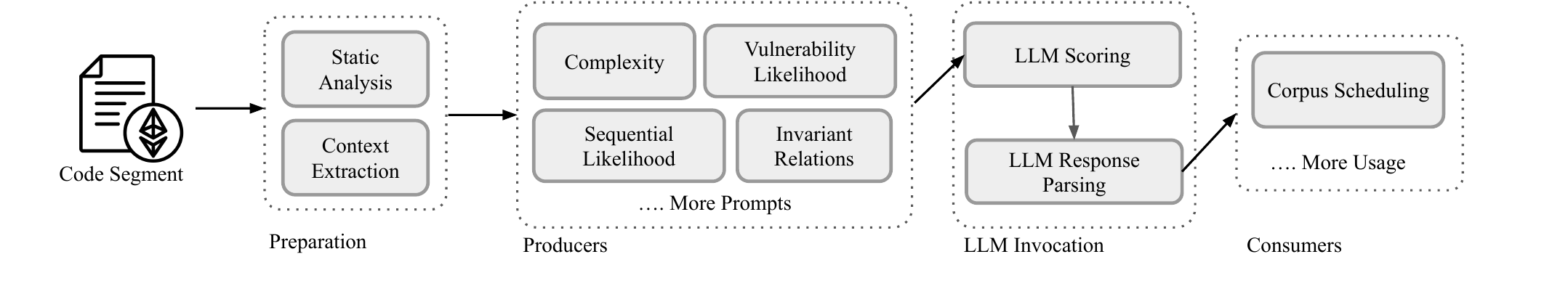}
  \caption{Workflow of \tool}
  \label{fig:workflow}
\end{figure*}

We present our techniques for using LLMs to guide and prioritize the fuzzing of smart contracts intelligently. 

\subsection{Overview}

The workflow of \tool\ is depicted in Figure~\ref{fig:workflow}.
\tool\ first converts each smart contract into its abstract syntax tree (AST) and then performs static analysis. This stage is important because we want œLLM to make well-informed decisions by providing necessary static analysis results. Then, the results are passed to producers, who extract scores using LLM from each code snippet created from the previous analysis. In the following sections, we introduce four producers: complexity, sequential likelihood, vulnerability likelihood, and invariant dependencies. The results of producers are encoded to be energy for the test cases in the corpus of \tool, which we collectively refer to as consumers, as it consumes the results from producers.  

\subsection{Parsing and Representation}

Beyond the source code of the smart contract, \tool{} provides additional contextual information for use with LLMs, enabling LLMs to make more accurate and reasonable decisions. Extracting these contexts is a complex task for the LLMs due to support for limited context lengths in LLMs, but it becomes more manageable through static analysis of smart contracts. Specifically, we extract cyclomatic complexity, which measures the number of independent paths through the source code; state variable dependencies, which provide insights into how different variables within the contract interact; external dependencies, which examine how the contract communicates with other external components; and control flow information, which offers a detailed look at the order in which the individual elements of the contract are executed. These attributes are carefully extracted based on methodologies and techniques developed in previous research.  The attributes are then used along with the source code to create the prompts. 

The task of submitting this detailed information to the language models presents its own set of challenges. The most straightforward approach would be to encapsulate all the information of the entire smart contract within a single prompt. However, most language models have a token limitation of 32K tokens, which creates a significant constraint, as almost all smart contract code exceeds this limit. To circumvent this limitation, we analyze the smart contract function by function, thereby reducing the overall size of the prompt. We begin by extracting all public functions within the smart contract, and for each function, we conduct a recursive analysis, identifying all related functions that a given function depends on by analyzing the call graph and data dependency graph. We then include the source code of these related functions in the prompt.

What distinguishes our approach is the reduction in prompt size and the careful prevention of information loss. By focusing only on the information that each function depends on with regard to calls, control flows, and dataflows, we could maintain the integrity of the information while conforming to the token limitations of the models. 

\subsection{Producer: Complexity}

The complexity of code segments within a program provides valuable insights into the challenges a fuzzer faces in covering each segment. By understanding this complexity, fuzzers can strategically allocate more time and resources to exploring the code's more intricate and challenging segments. This approach enhances the efficiency of fuzzing and ensures a more comprehensive examination of potential vulnerabilities.

Traditionally, this information is extracted using cyclomatic complexity. However, cyclomatic complexity is not an accurate representation of the complexity of the smart contract code. Firstly, cyclomatic complexity does not account for the statefulness of a smart contract. For instance, the following is the most complex function in an NFT smart contract, which transfers the NFT to another account. 
\begin{lstlisting}[language=Solidity]
function _transfer(address from, address to, uint256 tokenId) {
    require(ERC721.ownerOf(tokenId) == from, "ERC721: transfer from incorrect owner");
    _beforeTokenTransfer(from, to, tokenId);
    _approve(address(0), tokenId);
    _balances[from] -= 1;
    _balances[to] += 1;
    _owners[tokenId] = to;
    emit Transfer(from, to, tokenId);
    _afterTokenTransfer(from, to, tokenId);
}
\end{lstlisting}
The complexity mainly comes from the dependencies on \verb|_balances|, \verb|_owners|, etc., which are all state variables that can not be directly modified. To estimate the complexity of state dependencies, one needs to recursively lookup the functions writing to those state variables, which is extremely hard to find without false positives using static analysis or symbolic execution. LLMs, on the other hand, can reason with those stateful dependencies semantically, so unrelated state dependencies can be easily filtered based on the inference capability of LLMs. 

Next, cyclomatic complexity does not account for branch permissiveness (i.e., the likelihood of taking the branch with uniformly distributed inputs). For instance, in the code below, cyclomatic complexity concludes there is one branch, thus not complex. However, it is extremely complex to cover the true branch. Recent research~\cite{preach} proposes to use model counters to account for branch permissiveness. Yet, it takes a huge amount of computation resources and time.  
\begin{lstlisting}[language=Solidity]
if (pow(input, 2893) == 23947023413443923032) {  
    bug(); 
}
\end{lstlisting}

On the other hand, LLMs can understand the code semantically and see this as an extremely unpermissive branch, which implies the code is complex. 

\vspace{0.5cm}
\begin{tcolorbox}[colback=black!20!white, colframe=black!40!white]
\textbf{Human:} How complex is this code segment: \texttt{if (pow(input, 2893) == ...)} comparing to \texttt{if (pow(input, 2) == 9)}
\end{tcolorbox}

\begin{tcolorbox}[colback=white, colframe=black!20!white]
\textbf{GPT4:} The first code segment is more complex due to the large exponent, which could lead to numerical issues. The second code segment is simpler, raising the input to the power of 2.
\end{tcolorbox}
\vspace{0.5cm}

Lastly, smart contracts facilitate reentrancy (i.e., leaking control to the callers), hooks, and external calls. It is extremely hard to account for these cases using static analysis like cyclomatic complexity.  In our experience with LLMs, we found that LLMs can understand those types of transfer of controls and conclude complexity based on comments and code semantics. 

We construct a specialized prompt to analyze the complexity of functions in a smart contract using large language models. This prompt is designed to guide the models in evaluating complexity on a standardized scale of 100. By employing a uniform scale, we facilitate easy and meaningful comparisons between different functions within the contract.

The prompt is as follows:

\vspace{0.5cm}
\begin{tcolorbox}[colback=black!20!white, colframe=black!40!white]
How complex are the following Solidity code snippets (i.e., how hard is it to gain high test coverage by trying random arguments)? Rank within the range of 0 to 100. Output in the form of <Complexity 1>,<Complexity 2>,<Complexity 3>...
\end{tcolorbox}
\vspace{0.5cm}

The prompt is carefully structured to ensure the LLMs adhere to the specified format, eliminating unnecessary information. It is followed by the details of each function to be evaluated.

With this prompt, a language model replies with a list that denotes the complexity of each function. However, we recognize that a model may produce variations in the complexity values for the same function. To mitigate this inconsistency and derive a more reliable metric, we let the language model conduct the inference three times with different temperatures for each function, and then we compute the average of these three responses.

\subsection{Producer: Sequential Likelihood}

Existing fuzzing approaches for smart contracts like \textsc{ItyFuzz} and \textsc{Smartian} rely on static or dynamic analysis techniques to derive dependencies between contract functions and generate sequences of related functions as test cases. However, accurately tracking dataflow in smart contracts is challenging. Even when dataflow information is available, functions reading and writing the same state variables may still be independent in practice. For example, in the Uniswap pair contract, most functions write and access the ``lock'' variable simply to prevent reentrancy attacks, not because they are semantically related. Unlike static analysis, LLMs have more contextual knowledge about smart contract semantics. Thus, we leverage LLMs to generate interesting function call sequences that are likely to uncover new behaviors based on the following prompt.  

\vspace{0.5cm}
\begin{tcolorbox}[colback=black!20!white, colframe=black!40!white]
Suggest a series of interesting sequences given following the public Solidity functions and their code. Then, rank each interestingness of the sequence within the range of 0 to 100. Output one sequence in a line with the form of <Function Signature 1>=><Function Signature  2>:<Interestingness>. 
\end{tcolorbox}
\vspace{0.5cm}

\subsection{Other Producers}

Previous research\cite{tortoise}\cite{gf1} matches vulnerability patterns by similarity and allocates more resources to code regions prone to vulnerabilities. However, this method would fail if the vulnerable code is novel and not seen before or if additional code has been inserted around known vulnerable code segments. More recent work\cite{gf2}\cite{taint1}\cite{taint2} manually identifies a set of common vulnerability patterns and leverages taint analysis to locate potentially vulnerable code. Yet, it is inaccurate due to the statefulness of smart contracts, making it extremely hard to track dataflow. For LLMs, it can match the vulnerable code patterns logically and semantically. This information is ideal for supplementing fuzzing. For instance, a token contract may have a customized transfer function with buy tax, sell tax, and airdrop features. No accurate pattern can describe this transfer function's unintended or intended behavior. However, given the code segment of the function, LLMs can accurately summarize the logic and identify whether it may be susceptible to multiple attacks.  

To analyze the likelihood of vulnerabilities inside functions in a smart contract, we create the following prompt:

\vspace{0.5cm}
\begin{tcolorbox}[colback=black!20!white, colframe=black!40!white]
How likely are the following Solidity snippets to cause vulnerabilities (e.g., logical issue, reentrancy, etc.)? Rank each in terms of 100. Output in the form of <Likelihood 1>,<Likelihood 2>,<Likelihood 3>...

\end{tcolorbox}
\vspace{0.5cm}

We can also use the same method to analyze the invariants (user-defined assertions). Some commonly used invariants are that total assets in the smart contract should be greater after certain calls, and the value of a certain state variable shall never change. Specifically, given a set of invariants, we can ask LLMs to generate the likelihood of the code regions being a dependency of the invariant (i.e., to violate the invariant, how likely the code region must be exercised). 

\vspace{0.5cm}
\begin{tcolorbox}[colback=black!20!white, colframe=black!40!white]
How likely is following Solidity code snippets to cause \{\{Invariant\}\} being violated? Rank each in terms of 100. Output in the form of <Likelihood 1>,<Likelihood 2>,<Likelihood 3>...
\end{tcolorbox}
\vspace{0.5cm}


\subsection{Consumers: Fuzzing Prioritization}

We use a well-established corpus scheduling algorithm: power scheduling\cite{powerscheduling}\cite{powerschedule} to prioritize certain test cases based on complexity, vulnerability likelihood, state dependencies, or invariant dependencies. A test case with higher energy (i.e., $e(t)$) in the corpus is favored for more mutations during fuzzing. 


We created the following formula for complexity scores to derive the complexity energy (i.e., $K_{{\rm\it complexity}}(t)$, where $t$ is the test case). From LLM, we can generate a complexity score for each basic block. Intuitively, a test case is more interesting (i.e., having higher $K_{{\rm\it complexity}}(t)$) if its mutation can lead to more complex code region exploration as a different branch is taken (i.e., covering a neighboring basic block). Thus, for each basic block the test case can cover, \tool\ sums the complexity of their neighboring basic blocks to be the complexity score of the test case.  \tool\ then divides it by the total number of basic blocks covered by the test case to normalize the complexity scores. Following is the formula used to calculate $K_{{\rm\it complexity}}(t)$, where $C(b)$ denotes the complexity of the basic block $b$ generated by LLM, ${\rm\it BB}(t)$ returns all basic blocks executed by the test case $t$, and ${\rm\it neighbor}(b)$ is the set of neighboring basic blocks of the given basic block $b$ (i.e., basic blocks that share the same parent basic block). 
$$K_{{\rm\it complexity}}(t) = \frac{\sum_{b \in {\rm\it BB}(t)}{C({\rm\it neighbor}(b))}}{|{\rm\it BB}(t)|}$$
Invariant likelihood energy ($K_{{\rm \it invariant}}$) can also be derived similarly. 

During experiments, we identified that using a similar way to calculate vulnerability likelihood energy (i.e., $K_{{\rm \it vuln}}(t)$) leads to poor results. As shown in the motivating example, a vulnerable code region may need to be executed more than once in sequence to construct a state that could trigger a vulnerability. Using the previous way of calculation ignores the interestingness of repeating the same input again in the later part of the sequence.  Thus, for each test case, vulnerability likelihood energy is the sum of the vulnerability likelihood of all the basic blocks that got executed in the function. Below is the formula used to calculate $K_{{\rm \it vuln}}(t)$, where $V(b)$ denotes the vulnerability likelihood of the basic block $b$ generated by LLM, ${\rm\it Function}(t)$ returns all functions called by the test case $t$, and ${\rm\it BB}({\rm\it Function}(t))$ returns all the basic blocks in the functions called by the test case. 

$$K_{{\rm \it vuln}}(t) = \sum_{b \in {\rm\it BB}({\rm\it Function}(t))}{V(b)}$$

Sequential dependency is a producer providing completely different kinds of information. It provides a list of sequences of function calls with corresponding scores indicating how interesting the sequence is. For each test case, we extract all subsequences of the current sequence and sum the score of these subsequences.  The formula of sequential dependency energy (i.e., $K_{{\rm \it seq}}(t)$) is shown below, where function $S(s)$ returns the score of the given sequence (0 when the sequence is not present in the response returned by the LLM), and $2^{{\rm\it Seq}(t)}$ denotes all the subsequences of the test case. 
$$K_{{\rm \it seq}}(t) = \sum_{s \in 2^{{\rm\it Seq}(t)}}{S(s)}$$

To combine the energy derived from each producer ($K_{{\rm \it complexity}}, K_{{\rm \it invariant}}, K_{{\rm \it vuln}}, K_{{\rm \it seq}}$), we can use the following formula. In the formula, $e'(t)$ represents the original energy calculation method widely used by all fuzzers. To prevent the final energy from becoming too large, we cap it with $32$ times the original energy. This avoids assigning too much energy to a specific test case, preventing the possibility of exploring other code regions.  
$$e_{\rm\it cvs}(t) = {\rm\it min}(e'(t) + \sum_{K_i \in K}{e'(t) * K_i(t)},\ 32* e'(t))$$ 

In real-world experiments, we found that the fuzzer may not effectively utilize the complexity generated by an LLM.  LLMs provide complexity ranging from 0 to 100 and clusters around a specific value (i.e., having a small standard deviation). Thus, the energy calculated above also has a small standard deviation, diminishing the impact of complexity generated by LLMs. Instead, an ideal energy distribution shall have a large standard deviation (e.g., more than 100) to reflect the importance of code regions over others. 

To enhance the influence of this complexity, we adjust it using the following formula, where  $C_{{\rm\it fuzzer}}$ is the complexity provided to the fuzzer, and $C_{LLM}$ is the complexity generated by LLM.  $$C_{{\rm\it fuzzer}} = A ^ {C_{{\rm\it LLM}}} + B$$

$A$ and $B$ are constants obtained from the hyperparameter optimization of several campaigns. In our case, $A = 1.15$ and $B = 1200$ yield the best performance. Similarly, vulnerability likelihood shall be applied with the same technique before providing to the fuzzer, and similar $A$ and $B$ yield the best performance.

\section{Implementation}

We have implemented \tool\ on top of open-source \textsc{ItyFuzz} with 244 lines of Rust. We use Slither to conduct static analysis, analyze ASTs, and implement a preprocessor that queries LLM with 3k lines of Python.  For fast and cheap invocation of LLM, we selected the Llama 2 70B model\cite{llama} hosted at Anyscale and Replicate. We do not use GPT4 because it is too expensive. For each query, we invoke the LLM three times with temperatures 0.9, 0.95, and 1 and use the average of the results. We will open-source the tool upon acceptance. 

\section{Evaluation}
In this section, we attempt to answer the following research questions:
\begin{itemize}
    \item \noindent \textbf{RQ1 (Test Coverage)} Can \tool\ gain higher instruction coverage for various real-world projects? How many more instructions are covered by \tool?
    \item \noindent \textbf{RQ2 (Vulnerabilities)} Can \tool\ find more vulnerabilities in real-world projects than existing tools? Can \tool\ find new undiscovered vulnerabilities in well-audited projects?
    \item \noindent \textbf{RQ3 (Invariant)} Can \tool\ make invariant testing more efficient than existing property testing tools?
    \item \noindent \textbf{RQ4 (LLM Sensitivity)} Can different LLM producers provide unique and useful insights for \tool?
\end{itemize}

To answer these research questions, we collected 117 previously exploited smart contract projects, with LoC ranging from 100 to 100k, and compared the performance of \tool\ with \textsc{ItyFuzz}. \textsc{ItyFuzz} is selected as the baseline as it is the current state-of-the-art and has shown to be significantly better than all alternatives with respect to test coverage and the ability to detect real-world vulnerabilities. To minimize the human impact and negligence, when running \tool\ on vulnerable projects, we directly forked the chain at the immediate block before the exploit is executed. We also created multiple ablations of \tool. \tool-C (L-C) only uses complexity data for corpus scheduling.  \tool-V (L-V) only uses vulnerability likelihood data for corpus scheduling.  \tool-S (L-S) uses sequential likelihood data for state scheduling. 

In addition, we also run \tool\ on 600 unexploited on-chain projects on Binance Smart Chain, Base, Polygon, and Ethereum. 

All experiments were performed on a node with two Xeon E5-2698 V3 CPUs (64 threads) and 256 GB RAM. For each campaign, we repeat three times to minimize the impact of randomness in the fuzzer and report the average value. Unlike \textsc{ItyFuzz} paper, we run \tool\ on a single core to reduce the impact of IPC.

\subsection{Test Coverage}
\begin{figure}[htbp]
  \centering
  \includegraphics[scale=0.8]{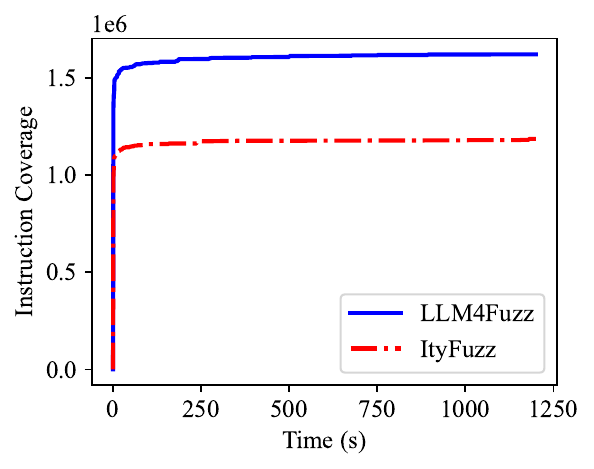}
  \caption{Total Instruction Coverage Over Time for Onchain Projects}
  \label{fig:cov_all}
\end{figure}

\begin{figure*}[htbp]
  \centering
  \includegraphics[scale=0.58]{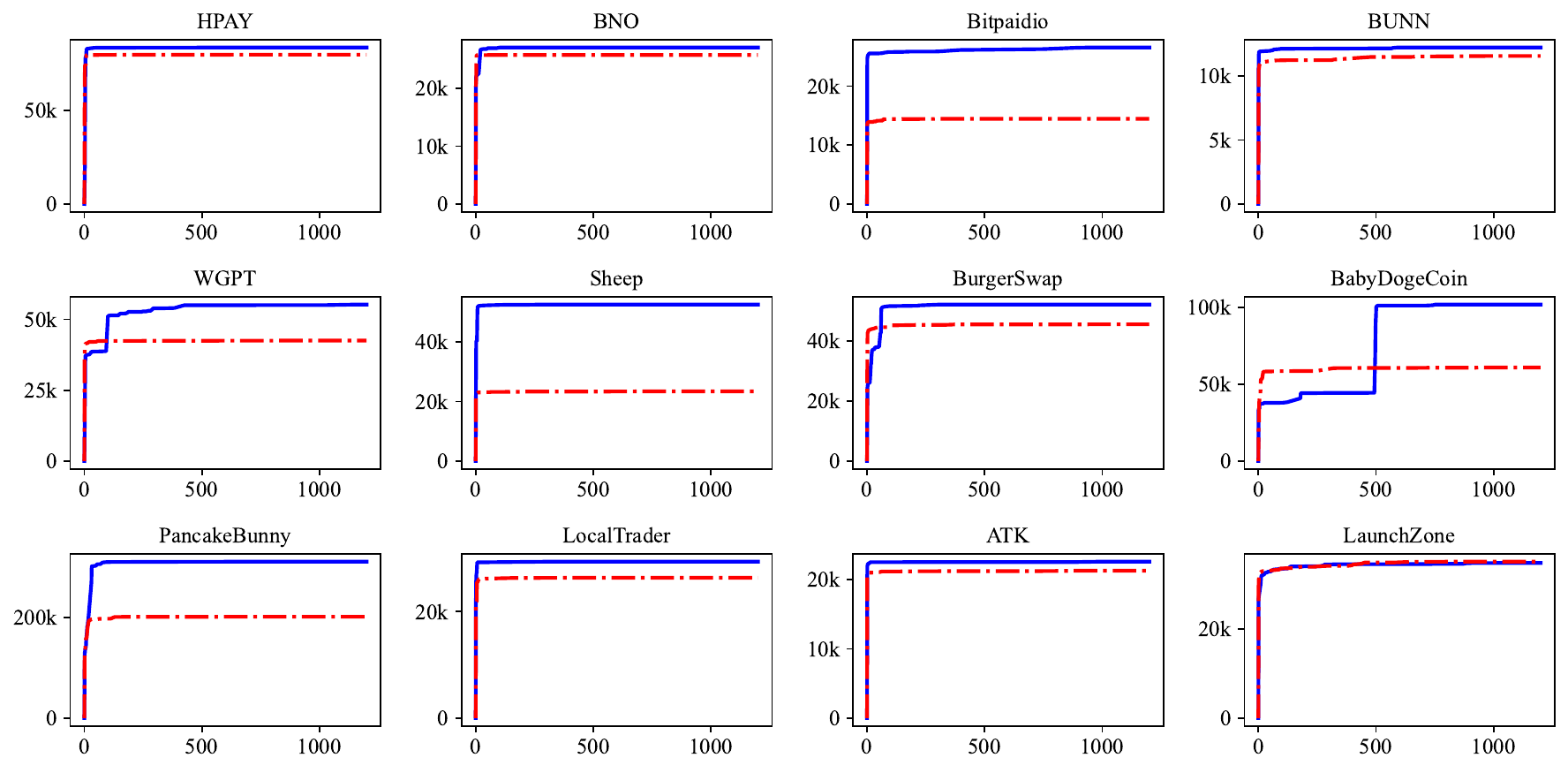}
  \caption{Test Coverage Over Time For Selected Onchain Projects (X is Instruction Coverage and Y is Time (s), Blue Concrete Line is \tool\ and Red Dotted Line is \textsc{ItyFuzz} (Baseline)}
  \label{fig:cov_individual}
\end{figure*}

In our assessment of 117 projects, the \tool\ could cover a total of 1.6M instructions\footnote{We evaluate instruction coverage as several previous research work~\cite{smartian}\cite{ityfuzz} have identified branch coverage is not a good representation of test coverage for smart contracts.}, in contrast to the baseline, which achieved coverage of only 1.1M instructions, as shown in Figure~\ref{fig:cov_all} and Figure~\ref{fig:increase}. We focused on a subset of 12 projects, with their individual coverage metrics presented in Figure~\ref{fig:cov_individual}. For specific projects, notably Bitpaidio and Sheep, the \tool\ realized a coverage that exceeded the baseline by more than two-fold. This large improvement can largely be attributed to the fact that, when informed by complexity, the fuzzer increasingly emphasizes more complex functions. 

Furthermore, the sequential likelihood metric is a valuable guide for the fuzzer. Traditionally, a fuzzer would require an average time complexity of $O(n^k)$ (where $n$ is the total number of functions, and $k$ represents the number of transactions required to access a specific code location) to navigate. Considering scenarios where $n$ can exceed 400 and $k$ might surpass 40, especially in real-world projects, the computational demands on the fuzzer can be prohibitively extensive. Discerning the optimal transaction sequences remains a formidable challenge despite integrating certain heuristics. Nevertheless, leveraging the sequential prioritization provided by LLMs, \tool\ can sidestep a significant portion of non-pertinent input sequences.

It is, however, imperative to acknowledge that the performance of the \tool\ is not universally superior. In certain projects, its efficacy is comparable to or even eclipsed by the baseline. Such deviations predominantly stem from instances where the LLM misinterprets specific code segments. A rigorous examination of instances where an LLM potentially mischaracterizes a smart contract is elucidated in \cite{gptscan}\cite{llmfails}.  


\subsection{Performance on Known Vulnerabilities}

\begin{table*}[h]
\centering
\begin{tabular}{|l|l|l|l|l|l|}
\hline
\textbf{Project Name} & \textbf{Time (Baseline)} & \textbf{Time (L-C)} & \textbf{Time (L-V)} & \textbf{Time (L-S)} & \textbf{Time (\tool)} \\
\hline
NewFreeDAO & inf & 4.9s \textbf{(-inf)} & 12.3s \textbf{(-inf)} & 161.7s \textbf{(-inf)} & 2.3s \textbf{(-inf)}\\ \hline
Axioma & 31.9s & 15.2s \textbf{(-16.7s)} & 4.0s \textbf{(-27.9s)} & 17.9s \textbf{(-14.0s)} & 1.2s \textbf{(-30.7s)}\\ \hline
FAPEN & 101.2s & 12.6s \textbf{(-88.6s)} & 75.6s \textbf{(-25.6s)} & 8.0s \textbf{(-93.2s)} & 16.9s \textbf{(-84.3s)}\\ \hline
cftoken & 15.4s & 1.9s \textbf{(-13.6s)} & 9.2s \textbf{(-6.2s)} & 4.4s \textbf{(-11.0s)} & 1.5s \textbf{(-13.9s)}\\ \hline
THB & 21.8s & inf \textbf{(inf)} & 18.2s \textbf{(-3.6s)} & 29.9s \textbf{(+8.1s)} & 247.0s \textbf{(+225.2s)}\\ \hline
RFB & inf & 14.7s \textbf{(-inf)} & 56.6s \textbf{(-inf)} & 88.1s \textbf{(-inf)} & 6.4s \textbf{(-inf)}\\ \hline
RES & 3.1s & inf \textbf{(+inf)} & inf \textbf{(+inf)} & 0.3s \textbf{(-2.7s)} & inf \textbf{(+inf)}\\ \hline
Yyds & 5.2s & 11.3s \textbf{(+6.0s)} & 12.8s \textbf{(+7.6s)} & 5.3s \textbf{(+0.1s)} & 11.1s \textbf{(+5.9s)}\\ \hline
Sheep & inf & 18.7s \textbf{(-inf)} & 92.7s \textbf{(-inf)} & inf & 12.5s \textbf{(-inf)}\\ \hline
AES & 6.6s & 3.0s \textbf{(-3.6s)} & 4.6s \textbf{(-2.1s)} & 2.8s \textbf{(-3.9s)} & 1.7s \textbf{(-4.9s)}\\ \hline
HEALTH & 62.3s & 9.6s \textbf{(-52.7s)} & 18.7s \textbf{(-43.7s)} & 40.1s \textbf{(-22.3s)} & 4.0s \textbf{(-58.3s)}\\ \hline
ApeDAO & 249.6s & 42.2s \textbf{(-207.3s)} & 1.6s \textbf{(-247.9s)} & 39.0s \textbf{(-210.6s)} & 2.1s \textbf{(-247.5s)}\\ \hline
BBOX & inf & 42.5s \textbf{(-inf)} & 0.8s \textbf{(-inf)} & 194.1s \textbf{(-inf)} & 1.5s \textbf{(-inf)}\\ \hline
ARA & inf & 52.1s \textbf{(-inf)} & 175.1s \textbf{(-inf)} & 957.8s \textbf{(-inf)} & 57.2s \textbf{(-inf)}\\ \hline
SEAMAN & 1112.3s & 72.0s \textbf{(-1040.3s)} & 12.3s \textbf{(-1100.1s)} & 8.8s \textbf{(-1103.6s)} & 3.9s \textbf{(-1108.4s)}\\ \hline
\end{tabular}
\caption{Vulnerability Detection Time for \tool, Ablations of \tool, and \textsc{ItyFuzz} (Baseline)}
\label{table:vulns}
\end{table*}

Among 117 projects, \textsc{ItyFuzz} can generate exploits for vulnerabilities in 45 projects within 259 seconds on average. \tool\, on the other hand, can generate exploits for vulnerabilities in 82 projects within 106 seconds on average. As shown in Table~\ref{table:vulns}, the vulnerability detection time for the \tool\ and its ablations (\tool-C, \tool-V, \tool-S) is overall better compared to the baseline (\textsc{ItyFuzz}) among 15 randomly selected projects. While several projects, including NewFreeDAO, RFB, Sheep, BBOX, and ARA, demonstrate an infinite detection time for the baseline, all \tool\ variations present finite and often vastly reduced times, showing the efficacy of \tool\ even in scenarios where the baseline falters entirely.

Among the ablations, \tool-C has the shortest time finding vulnerabilities for most projects. \tool\ also performs consistently better than any of the ablations. This indicates that a combination of the producers has reduced detection time than what can be reduced by a single producer.  


For certain projects, like RES, \tool\ performs worse or even completely unable to detect vulnerabilities compared to the baseline approach. From ablations, we can clearly see that when \tool\ performs worse, \tool-C or \tool-V also perform worse. Thus, it is likely because LLM provided a low complexity or low vulnerability likelihood score for critical functions to trigger the vulnerabilities. As discussed previously, it is inevitable for LLM to misunderstand certain parts of the contracts. One possible solution would be using a more complex but expensive LLM like GPT-4.

\subsection{New Vulnerabilities Found}

In addition to the vulnerabilities listed in Table~\ref{table:vulns}, we also found \textbf{5 projects having critical vulnerabilities} among 600 projects that 
have gone through audits with reputable auditing firms. The affected projects hold a total \$247k worth of assets. For those five critical vulnerabilities, existing tools like Mythril and Echidna can find none of them. \textsc{ItyFuzz} can find 3 of them, with one uncovered in 30 minutes and two taking days. On the other hand, \tool\ can uncover all of them in less than 30 minutes.

For instance, \tool{} found a project comprising an ERC20 token, multiple liquidity pools, and governance deployed on Binance Smart Chain to be exploitable. The vulnerability arises from two minor logical bugs that, on their own, are not exploitable.

The first small bug is in the implementation of the \verb|_transfer| function (Line 1 - 12 in Figure~\ref{fig:vul_sc}). Specifically, \verb|fee| amount of token is minted (i.e., created and transferred) to the token contract when the user transfers tokens to the liquidity pool (i.e., selling the token). Given that the \verb|fee| is multiplied by 2, if the attackers transfer multiple times to the liquidity pool (and have the liquidity pool refund them), they can make the sum of their balance and the contract significantly greater than they initially have. This small bug increases the total supply of this token, and the bug itself cannot be exploited to steal funds.

The second bug exists in function \verb|fund| (Line 13-27). The intention of this function is to reward owners with the \verb|fee| collected in the contract.  Instead of rewarding the owners with that token, it swaps those tokens to USDT (a token that can directly be redeemed to US Dollars) and transfers them to the owners. During the swap process, the minimum USDT to receive is incorrectly set to 0, which allows sandwich attacks (the attacker sells a lot of tokens, then the victim sells the token but receives almost nothing, and finally, the attacker buys back the token at a lower price, profiting from the selling). These two bugs combined can lead to critical vulnerabilities, allowing attackers to steal funds. Specifically, the attacker can leverage the first bug to mint a lot of tokens in the contract (e.g., to mint $v$ amount of tokens in the contract, the attacker only needs to spend $v/2$ tokens) and then conduct sandwich attacks to extract these tokens.  

\begin{figure}
\begin{lstlisting}[language=Solidity]
function _transfer(address from, address to, uint256 amount) internal override {
    ...
    if (isAMM(from, to) && !isAL(from, to)) { 
        fee = amount * fs / b;
        realAmount = amount - fee;
        super._mint(address(this), fee * 2);
        super._burn(from, fee);
        super._transfer(from, to, realAmount);
        
    }
    ...
}
function fund() public {
    address[] memory path = new address[](2);
    uint256 ta = IERC20(address(this)).balanceOf(address(this));
    path[0] = address(this);
    path[1] = USDT;
    uniswapV2Router.swapExactTokensForTokens(
        ta / 2, 0, path,
        owner1, block.timestamp
    );
    uniswapV2Router.swapExactTokensForTokens(
        ta / 2, 0,  path,
        owner2, block.timestamp
    );
    ...
}
\end{lstlisting}
\label{fig:vul_sc}
\caption{Snippet of the vulnerable smart contract.\label{fig:vulnerable}}
\end{figure}

However, it is non-trivial to leverage these two bugs to trigger the vulnerability. Notice that deflating tokens also leads to the fee being taken. It is hard to make a profitable attack without intricate calculations of how many transfers are required and how many tokens to borrow. \textsc{ItyFuzz} cannot find a profitable exploit in 24 hours. On the other hand, \tool\ can identify a profitable exploit consisting of at least 28 transactions in 1.9 hours on average. Specifically, the LLM has assigned high complexity and high vulnerability likelihood to \verb|transfer|, \verb|transferFrom|, and \verb|fund|. In addition, the sequential likelihood for \verb|transfer|, \verb|skim| (having liquidity pool refund), \verb|fund| is the highest among all other sequences. 

\subsection{Invariant Violation Detection Efficiency}

To answer RQ3, we manually crafted four invariants that can be violated under extreme conditions (i.e., cannot be violated with simple invariant testing tools like Echidna\cite{echidna}). We show the results in Table~\ref{invariants}. While \textsc{ItyFuzz} takes minutes or times out to find invariant violations, most of the time, \tool\ can find the violations in less than a minute. By examining traces of \textsc{ItyFuzz}, we identify that \textsc{ItyFuzz} commonly got stuck in certain functions that are not at all related to the violations. For instance, in Uniswap V2, \textsc{ItyFuzz} spends a significant amount of time exploring \verb|mint| and \verb|burn|, while \tool\ spends most of the time on finding the correct \verb|swap| transaction, which is necessary to execute before triggering the invariant violation. 

\begin{table}[h]
\centering
\begin{tabular}{|l|l|l|}
\hline
\textbf{Invariant Name} & \textbf{\textsc{ItyFuzz}} & \textbf{\tool} \\
\hline
Uniswap V2 Flashloan & 459.1s & 14.2s \\
\hline
Uniswap V3 Flashloan & Timeout & 92.0s \\
\hline
Team Finance & 139.4s & 6.5s \\
\hline
Sublime Pool Miscalculation & 190.1s & 58.8s \\
\hline
\end{tabular}

\label{invariants}

\caption{Invariants Violation Time for \textsc{ItyFuzz} and \tool}
\end{table}

We also tried \tool\ on Daedaluzz, a benchmark for smart contract property testing tools. However, there is no significant improvement because the invariants are manually inserted into a generated puzzle. It is hard for LLM to extract complexity or invariants scoring information. By intuition, if LLM can extract this information on code without semantics, then it can be a constant-time replacement of constraint solver or model counter, which is impossible. However, we recognize that LLM-guided concolic execution for hybrid fuzzing may help. We yield this to future work.

\subsection{Impact of the LLM}

\begin{figure}[htbp]
  \centering
  \includegraphics[scale=0.85]{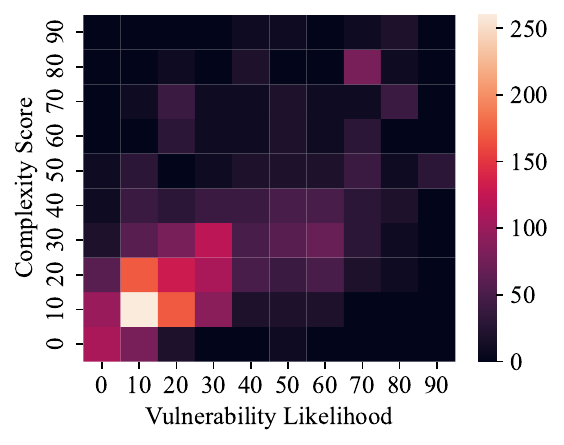}
  \caption{Amount of Code Regions with Certain Complexity and Vulnerability Likelihood}
  \label{fig:complexity-vulns}
\end{figure}

\begin{figure}[htbp]
  \centering
  \includegraphics[scale=0.85]{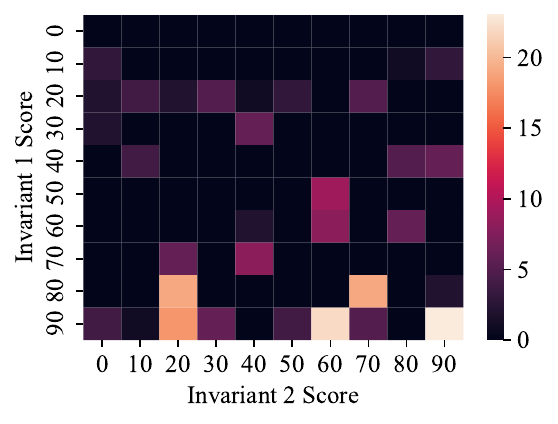}
  \caption{Amount of Code Regions with Certain Invariant 1 and 2 Dependency Score}
  \label{fig:invariants-comp}
\end{figure}

Some may consider that LLMs cannot analyze programs; that is, LLMs generate the same dummy metrics for different producers in the same code regions. We show this claim is incorrect by contrasting the complexity and vulnerability likelihood generated by the LLM for each code region. In Figure~\ref{fig:complexity-vulns}, a heatmap showing the number of code regions that have certain complexity scores and vulnerability likelihood, we demonstrate that complexity score and vulnerability likelihood for the same code regions are mostly different (i.e., not forming a straight line on the heatmap). Though there are lots of code regions having both complexity and vulnerability likelihood to be 10 (expected as some code regions do nothing, so they are not complex and not vulnerable), generally, these two metrics are weakly correlated, especially for code regions having complexity or vulnerability likelihood to be greater than 10. These observations imply that the LLM can distinguish these two producers without providing dummy metrics. In Figure~\ref{fig:invariants-comp}, we further demonstrate that the LLM can generate different metrics regarding different invariants on the same code regions. This confirms that the LLM can also distinguish between invariants.


\subsection{Cost and Time}
We used Llama 2 70B for all queries. The cost ranges from \$1 per 1M tokens (Anyscale) to >\$3 per 1M tokens (Replicate). On average, \tool\ spends \$0.15 for each project and takes 19 seconds for all queries to be responded to. 

\subsection{Extending to Traditional Software}
\begin{figure*}
    \centering
    \subfigure[coverage of readelf]{
        \includegraphics[width=0.4\linewidth]{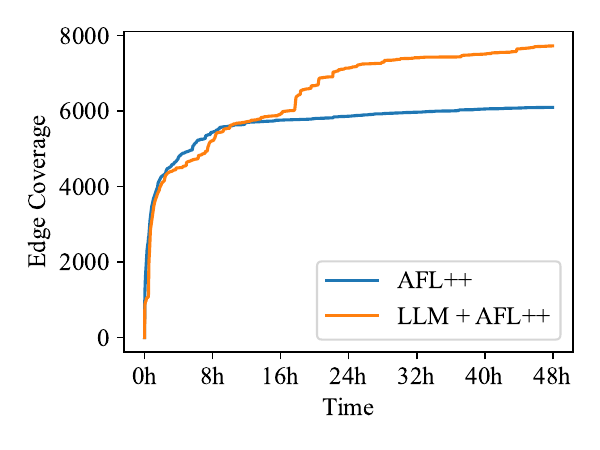}
        \label{subfig:cov-readelf}
    }
    \subfigure[coverage of nm-new]{
        \includegraphics[width=0.4\linewidth]{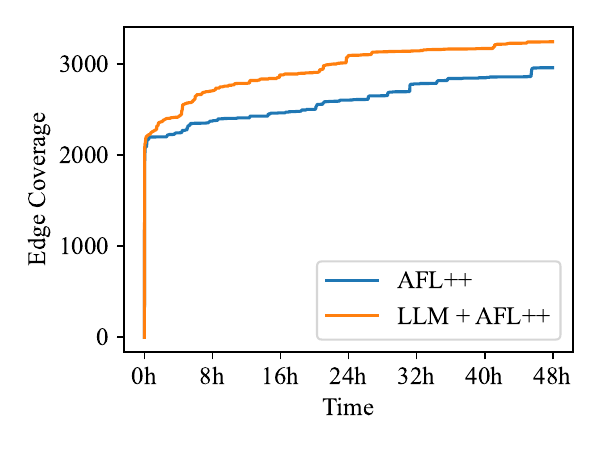}
        \label{subfig:cov-nm-new}
    }
    \caption{\texttt{binutils} coverage}
    \label{fig:cov-binutils}
    \vspace{-0.4cm}
\end{figure*}

To demonstrate the ideas in \tool\ are also feasible for general software, we have adapted algorithms of \tool\ for AFL++\cite{aflpp}. The scheduler biases AFL++ towards test cases with higher scores, promoting the exploration of more complex and potentially vulnerable code paths identified by the LLM. Figure~\ref{fig:cov-binutils} demonstrates that AFL++ with LLM guidance gains more coverage than simply using AFL++ on binutils programs. Future work can conduct more research on this topic. 

\section{Related Work}

\paragraph{Smart Contract Testing.}
Prior work like Echidna \cite{echidna} ILF \cite{ilf} and Harvey \cite{harvey} are fuzzers that find vulnerabilities triggered by a single or a sequence of inputs. They are guided by test coverage or other static, dynamic analysis. \textsc{Smartian}\cite{smartian} and \textsc{ItyFuzz}\cite{ityfuzz} extends these prior work by using dataflow and comparisons to schedule sequential mutations efficiently. However, accurate dataflow tracking scales poorly, and programs often have semantically meaningless dataflow. 

There are also concolic testing tools like Mythril \cite{mythril} and static analysis tools like Slither \cite{slither}. Yet, none of them is as performant as fuzzers like \textsc{ItyFuzz} and \textsc{Smartian} as demonstrated in their evaluations. Recent work\cite{gptscan} combines LLMs with static analysis for smart contracts, leveraging static analysis to filter LLM vulnerability reports for false positives. Compared to fuzzing, it has higher false negatives and has trouble detecting more subtle logical vulnerabilities that could lead to fund loss. 

\paragraph{LLM and Fuzzing.}

Recent research activities have primarily been focusing on using LLM to generate a corpus for fuzzing \cite{llm2}\cite{llm5}\cite{llm6}.  They let an LLM generate interesting test cases and provide them as seeds for the fuzzer before the fuzzer starts or during the runtime. Deng et al.\cite{llm3}\cite{llm4} and Xia et al.\cite{llm7} have also treated LLM as a fuzzer. They instruct LLM to generate structural inputs with respect to programs and mutate them during fuzzing. They observe that an LLM itself is a structural fuzzer that can outcompete traditional structural fuzzers that generate inputs using grammar. Other research work\cite{llminv}\cite{llm1}\cite{googlellm} also leverages LLM to generate fuzzer drivers and invariants that allow fuzzers to identify deeper logical issues. Lastly, LLMs have been widely used in program debugging and repair\cite{debug1}\cite{debug2}\cite{repair1}\cite{repair2}\cite{repair3}. Yet, none has integrated an LLM with a fuzzer like \tool\ to create a prioritization scheme to facilitate fuzzer exploration. 

\paragraph{Stateful Fuzzing.}

\textsc{ItyFuzz}\cite{ityfuzz} and \textsc{Smartian}\cite{smartian} as described previously are complex stateful fuzzers. In traditional software testing, multiple stateful fuzzers\cite{sgf}\cite{nyx}\cite{nyxnet}\cite{dongdong}\cite{ijon} have been proposed. For instance, IJON\cite{ijon} exposes the state to guide fuzzing towards new program states. \textsc{Nyx}\cite{nyx} employs a rapid snapshotting approach at the operating system level, while its extension, \textsc{Nyx-Net}\cite{nyxnet}, utilizes these snapshotting methods to revert to prior states for efficient stateful fuzzing for network applications. Comparably, Dong et al.\cite{dongdong} leverage incremental snapshotting of the Android OS when fuzzing Android applications so that the fuzzer can travel back to a previous application state. However, none of these stateful fuzzers have complex stateful guidance as in \textsc{ItyFuzz} or \textsc{Smartian} due to the cost and complexity of tracking runtime information. Techniques proposed in \tool\ can potentially be integrated into all those stateful fuzzers to reduce overhead on exploring uninteresting states or sequences of inputs.  

\paragraph{Guided Fuzzing.}


MTFuzz\cite{mtfuzz} uses multi-task neural networks to compact input embeddings across diverse tasks, enabling the fuzzer to focus mutations on influential bytes that can increase coverage. K-Scheduler\cite{katz} improves efficiency by building an edge horizon graph from the CFG and using Katz centrality to prioritize seeds reaching more unvisited edges. GreyOne\cite{greyone} leverages lightweight fuzzing-driven taint inference to guide mutations towards input bytes affecting more untouched branches. \tool\ can be used in conjunction with these techniques. 

\section{Discussions}

\paragraph{Additional Consumers.} We recognize that an additional consumer of the LLM-generated complexity and vulnerability likelihood could be concolic execution. By solving branches that are more complex and likely to lead to vulnerabilities, there could be more synergy between fuzzers in charge of exploring simple program paths and concolic execution exploring complex program paths.


\paragraph{Fine Tuning.} Fine-tuning the LLMs on smart contract code and known vulnerabilities would enhance their specialized knowledge in this domain. With greater exposure to Solidity programming patterns, historical exploits, execution traces, and other smart contract data, the model could build a more comprehensive understanding of code semantics, security flaws, and state transitions. This would empower LLMs to conduct more nuanced analysis when guiding a fuzzer.

\paragraph{Consensus Reaching Among LLMs.} In our experiment, we only used one LLM. We recognize there are multiple LLMs available that can give an accurate response to queries from \tool. By reaching a consensus (e.g., averaging response) among multiple LLMs, the producers can likely generate a value that is less likely to be biased.   

\paragraph{Small Variations on Prompt.} We have not attempted to find the optimized prompts for each producer. Future work can leverage techniques like \cite{prompt} to create a better prompt.

\section{Ethical Consideration}

We have disclosed all vulnerabilities to project owners. 
Due to the nature of blockchain, it is impossible to update the smart contract code once deployed. 
Project owners are encouraged to white hat hack their projects. 


\section{Conclusion}

This paper presents \tool, a methodology that leverages large language models to guide and prioritize the fuzzing of smart contracts. By extracting code attributes and crafting specialized prompts, \tool\ produces metrics for code complexity, vulnerability likelihood, invariant relations, and input sequences. These metrics are encoded into the corpus scheduler of the fuzzer to focus testing on more valuable code regions. Evaluations show \tool\ achieves substantially higher coverage and detects more vulnerabilities than state-of-the-art tools, including finding critical flaws in live contracts. \tool\ overcomes traditional fuzzing limitations by harnessing LLMs' semantic reasoning power to explore smart contracts more efficiently. As blockchain adoption increases, LLM-guided fuzzing can provide impactful assistance in securing smart contracts.

\bibliographystyle{plain}
\bibliography{main}

\begin{thebibliography}{10}

\bibitem{aes}
{AES} {Attack} {Explanation}.
\newblock \url{https://twitter.com/BlockSecTeam/status/1600442137811689473}.

\bibitem{googlellm}
{AI}-{Powered} {Fuzzing}: {Breaking} the {Bug} {Hunting} {Barrier}.
\newblock
  \url{https://security.googleblog.com/2023/08/ai-powered-fuzzing-breaking-bug-hunting.html}.

\bibitem{tvl2}
All {Chains} {TVL}.
\newblock \url{https://defillama.com/chains}.

\bibitem{erc20}
{ERC}-20 {Token} {Standard}.
\newblock
  \url{https://ethereum.org/en/developers/docs/standards/tokens/erc-20/}.

\bibitem{gpt4}
{GPT}-4.
\newblock \url{https://openai.com/research/gpt-4}.

\bibitem{llama}
Llama 2 {70B} {Model}.
\newblock \url{https://ai.meta.com/llama/}.

\bibitem{powerscheduling}
Power {Schedules}.
\newblock \url{https://aflplus.plus/docs/power_schedules/}.

\bibitem{uniswapv2}
Uniswap-v2 {Contract} {Walk}-{Through}.
\newblock
  \url{https://ethereum.org/en/developers/tutorials/uniswap-v2-annotated-code/}.

\bibitem{llm6}
Joshua Ackerman and George Cybenko.
\newblock Large language models for fuzzing parsers (registered report).
\newblock In {\em Proceedings of the 2nd International Fuzzing Workshop}, pages
  31--38, 2023.

\bibitem{ijon}
Cornelius Aschermann, Sergej Schumilo, Ali Abbasi, and Thorsten Holz.
\newblock Ijon: Exploring deep state spaces via fuzzing.
\newblock In {\em 2020 IEEE Symposium on Security and Privacy (SP)}, pages
  1597--1612. IEEE, 2020.

\bibitem{attacksurvey}
Nicola Atzei, Massimo Bartoletti, and Tiziana Cimoli.
\newblock A survey of attacks on ethereum smart contracts (sok).
\newblock In {\em Principles of Security and Trust: 6th International
  Conference, POST 2017, Held as Part of the European Joint Conferences on
  Theory and Practice of Software, ETAPS 2017, Uppsala, Sweden, April 22-29,
  2017, Proceedings 6}, pages 164--186. Springer, 2017.

\bibitem{sgf}
Jinsheng Ba, Marcel B{\"o}hme, Zahra Mirzamomen, and Abhik Roychoudhury.
\newblock Stateful greybox fuzzing.
\newblock In {\em 31st USENIX Security Symposium (USENIX Security 22)}, pages
  3255--3272, 2022.

\bibitem{powerschedule}
Marcel B\"{o}hme, Van-Thuan Pham, and Abhik Roychoudhury.
\newblock Coverage-based greybox fuzzing as markov chain.
\newblock In {\em Proceedings of the 2016 ACM SIGSAC Conference on Computer and
  Communications Security}, CCS '16, page 1032–1043, New York, NY, USA, 2016.
  Association for Computing Machinery.

\bibitem{llmok}
Chong Chen, Jianzhong Su, Jiachi Chen, Yanlin Wang, Tingting Bi, Yanli Wang,
  Xingwei Lin, Ting Chen, and Zibin Zheng.
\newblock When chatgpt meets smart contract vulnerability detection: How far
  are we?
\newblock {\em arXiv preprint arXiv:2309.05520}, 2023.

\bibitem{smartian}
Jaeseung Choi, Doyeon Kim, Soomin Kim, Gustavo Grieco, Alex Groce, and Sang~Kil
  Cha.
\newblock Smartian: Enhancing smart contract fuzzing with static and dynamic
  data-flow analyses.
\newblock In {\em 2021 36th IEEE/ACM International Conference on Automated
  Software Engineering (ASE)}, pages 227--239. IEEE, 2021.

\bibitem{llmfails}
Isaac David, Liyi Zhou, Kaihua Qin, Dawn Song, Lorenzo Cavallaro, and Arthur
  Gervais.
\newblock Do you still need a manual smart contract audit?
\newblock {\em arXiv preprint arXiv:2306.12338}, 2023.

\bibitem{llm3}
Yinlin Deng, Chunqiu~Steven Xia, Haoran Peng, Chenyuan Yang, and Lingming
  Zhang.
\newblock Fuzzing deep-learning libraries via large language models.
\newblock {\em arXiv preprint arXiv:2212.14834}, 2022.

\bibitem{llm4}
Yinlin Deng, Chunqiu~Steven Xia, Haoran Peng, Chenyuan Yang, and Lingming
  Zhang.
\newblock Large language models are zero-shot fuzzers: Fuzzing deep-learning
  libraries via large language models.
\newblock In {\em Proceedings of the 32nd ACM SIGSOFT International Symposium
  on Software Testing and Analysis}, pages 423--435, 2023.

\bibitem{survey2}
Monika Di~Angelo and Gernot Salzer.
\newblock A survey of tools for analyzing ethereum smart contracts.
\newblock In {\em 2019 IEEE international conference on decentralized
  applications and infrastructures (DAPPCON)}, pages 69--78. IEEE, 2019.

\bibitem{mythril}
Monika Di~Angelo and Gernot Salzer.
\newblock A survey of tools for analyzing ethereum smart contracts.
\newblock In {\em 2019 IEEE international conference on decentralized
  applications and infrastructures (DAPPCON)}, pages 69--78. IEEE, 2019.

\bibitem{dongdong}
Zhen Dong, Marcel B{\"o}hme, Lucia Cojocaru, and Abhik Roychoudhury.
\newblock Time-travel testing of android apps.
\newblock In {\em Proceedings of the ACM/IEEE 42nd International Conference on
  Software Engineering}, pages 481--492, 2020.

\bibitem{debug1}
Zachary Englhardt, Richard Li, Dilini Nissanka, Zhihan Zhang, Girish
  Narayanswamy, Joseph Breda, Xin Liu, Shwetak Patel, and Vikram Iyer.
\newblock Exploring and characterizing large language models for embedded
  system development and debugging.
\newblock {\em arXiv preprint arXiv:2307.03817}, 2023.

\bibitem{slither}
Josselin Feist, Gustavo Grieco, and Alex Groce.
\newblock Slither: a static analysis framework for smart contracts.
\newblock In {\em 2019 IEEE/ACM 2nd International Workshop on Emerging Trends
  in Software Engineering for Blockchain (WETSEB)}, pages 8--15. IEEE, 2019.

\bibitem{aflpp}
Andrea Fioraldi, Dominik Maier, Heiko Ei{\ss}feldt, and Marc Heuse.
\newblock $\{$AFL++$\}$: Combining incremental steps of fuzzing research.
\newblock In {\em 14th USENIX Workshop on Offensive Technologies (WOOT 20)},
  2020.

\bibitem{greyone}
Shuitao Gan, Chao Zhang, Peng Chen, Bodong Zhao, Xiaojun Qin, Dong Wu, and
  Zuoning Chen.
\newblock $\{$GREYONE$\}$: Data flow sensitive fuzzing.
\newblock In {\em 29th USENIX security symposium (USENIX Security 20)}, pages
  2577--2594, 2020.

\bibitem{echidna}
Gustavo Grieco, Will Song, Artur Cygan, Josselin Feist, and Alex Groce.
\newblock Echidna: effective, usable, and fast fuzzing for smart contracts.
\newblock In {\em Proceedings of the 29th ACM SIGSOFT International Symposium
  on Software Testing and Analysis}, pages 557--560, 2020.

\bibitem{llm2}
Qiuhan Gu.
\newblock Llm-based code generation method for golang compiler testing.
\newblock 2023.

\bibitem{ilf}
Jingxuan He, Mislav Balunovi{\'c}, Nodar Ambroladze, Petar Tsankov, and Martin
  Vechev.
\newblock Learning to fuzz from symbolic execution with application to smart
  contracts.
\newblock In {\em Proceedings of the 2019 ACM SIGSAC Conference on Computer and
  Communications Security}, pages 531--548, 2019.

\bibitem{llm5}
Jie Hu, Qian Zhang, and Heng Yin.
\newblock Augmenting greybox fuzzing with generative ai.
\newblock {\em arXiv preprint arXiv:2306.06782}, 2023.

\bibitem{repair2}
Matthew Jin, Syed Shahriar, Michele Tufano, Xin Shi, Shuai Lu, Neel Sundaresan,
  and Alexey Svyatkovskiy.
\newblock Inferfix: End-to-end program repair with llms.
\newblock {\em arXiv preprint arXiv:2303.07263}, 2023.

\bibitem{llminv}
Rahul Kande, Hammond Pearce, Benjamin Tan, Brendan Dolan-Gavitt, Shailja
  Thakur, Ramesh Karri, and Jeyavijayan Rajendran.
\newblock Llm-assisted generation of hardware assertions.
\newblock {\em arXiv preprint arXiv:2306.14027}, 2023.

\bibitem{repair3}
Sungmin Kang, Juyeon Yoon, and Shin Yoo.
\newblock Large language models are few-shot testers: Exploring llm-based
  general bug reproduction.
\newblock In {\em 2023 IEEE/ACM 45th International Conference on Software
  Engineering (ICSE)}, pages 2312--2323. IEEE, 2023.

\bibitem{debug2}
Haonan Li, Yu~Hao, Yizhuo Zhai, and Zhiyun Qian.
\newblock The hitchhiker's guide to program analysis: A journey with large
  language models.
\newblock {\em arXiv preprint arXiv:2308.00245}, 2023.

\bibitem{taint2}
Yuwei Li, Shouling Ji, Chenyang Lv, Yuan Chen, Jianhai Chen, Qinchen Gu, and
  Chunming Wu.
\newblock V-fuzz: Vulnerability-oriented evolutionary fuzzing.
\newblock {\em arXiv preprint arXiv:1901.01142}, 2019.

\bibitem{blockchaingrowth}
Ahmed~Afif Monrat, Olov Schel{\'e}n, and Karl Andersson.
\newblock A survey of blockchain from the perspectives of applications,
  challenges, and opportunities.
\newblock {\em IEEE Access}, 7:117134--117151, 2019.

\bibitem{gf2}
Sebastian {\"O}sterlund, Kaveh Razavi, Herbert Bos, and Cristiano Giuffrida.
\newblock $\{$ParmeSan$\}$: Sanitizer-guided greybox fuzzing.
\newblock In {\em 29th USENIX Security Symposium (USENIX Security 20)}, pages
  2289--2306, 2020.

\bibitem{fuzzfactory}
Rohan Padhye, Caroline Lemieux, Koushik Sen, Laurent Simon, and Hayawardh
  Vijayakumar.
\newblock Fuzzfactory: domain-specific fuzzing with waypoints.
\newblock {\em Proceedings of the ACM on Programming Languages},
  3(OOPSLA):1--29, 2019.

\bibitem{preach}
Seemanta Saha, Laboni Sarker, Md~Shafiuzzaman, Chaofan Shou, Albert Li, Ganesh
  Sankaran, and Tevfik Bultan.
\newblock Rare path guided fuzzing.
\newblock In {\em Proceedings of the 32nd ACM SIGSOFT International Symposium
  on Software Testing and Analysis}, ISSTA 2023, page 1295–1306, New York,
  NY, USA, 2023. Association for Computing Machinery.

\bibitem{nyx}
Sergej Schumilo, Cornelius Aschermann, Ali Abbasi, Simon W{\"o}rner, and
  Thorsten Holz.
\newblock Nyx: Greybox hypervisor fuzzing using fast snapshots and affine
  types.
\newblock In {\em 30th USENIX Security Symposium (USENIX Security 21)}, pages
  2597--2614, 2021.

\bibitem{nyxnet}
Sergej Schumilo, Cornelius Aschermann, Andrea Jemmett, Ali Abbasi, and Thorsten
  Holz.
\newblock Nyx-net: network fuzzing with incremental snapshots.
\newblock In {\em Proceedings of the Seventeenth European Conference on
  Computer Systems}, pages 166--180, 2022.

\bibitem{mtfuzz}
Dongdong She, Rahul Krishna, Lu~Yan, Suman Jana, and Baishakhi Ray.
\newblock Mtfuzz: fuzzing with a multi-task neural network.
\newblock In {\em Proceedings of the 28th ACM joint meeting on European
  software engineering conference and symposium on the foundations of software
  engineering}, pages 737--749, 2020.

\bibitem{katz}
Dongdong She, Abhishek Shah, and Suman Jana.
\newblock Effective seed scheduling for fuzzing with graph centrality analysis.
\newblock In {\em 2022 IEEE Symposium on Security and Privacy (SP)}, pages
  2194--2211. IEEE, 2022.

\bibitem{ityfuzz}
Chaofan Shou, Shangyin Tan, and Koushik Sen.
\newblock Ityfuzz: Snapshot-based fuzzer for smart contract.
\newblock In {\em Proceedings of the 32nd ACM SIGSOFT International Symposium
  on Software Testing and Analysis}, pages 322--333, 2023.

\bibitem{smartest}
Sunbeom So, Seongjoon Hong, and Hakjoo Oh.
\newblock $\{$SmarTest$\}$: Effectively hunting vulnerable transaction
  sequences in smart contracts through language $\{$Model-Guided$\}$ symbolic
  execution.
\newblock In {\em 30th USENIX Security Symposium (USENIX Security 21)}, pages
  1361--1378, 2021.

\bibitem{tvl1}
Viktorija Stepanova and Ingars Eri{\c{n}}{\v{s}}.
\newblock Review of decentralized finance applications and their total value
  locked.
\newblock {\em TEM Journal}, 10(1), 2021.

\bibitem{llmok2}
Andr{\'e} Storhaug, Jingyue Li, and Tianyuan Hu.
\newblock Efficient avoidance of vulnerabilities in auto-completed smart
  contract code using vulnerability-constrained decoding.
\newblock {\em arXiv preprint arXiv:2309.09826}, 2023.

\bibitem{gptscan}
Yuqiang Sun, Daoyuan Wu, Yue Xue, Han Liu, Haijun Wang, Zhengzi Xu, Xiaofei
  Xie, and Yang Liu.
\newblock When gpt meets program analysis: Towards intelligent detection of
  smart contract logic vulnerabilities in gptscan.
\newblock {\em arXiv preprint arXiv:2308.03314}, 2023.

\bibitem{survey3}
Palina Tolmach, Yi~Li, Shang-Wei Lin, Yang Liu, and Zengxiang Li.
\newblock A survey of smart contract formal specification and verification.
\newblock {\em ACM Computing Surveys (CSUR)}, 54(7):1--38, 2021.

\bibitem{survey1}
Anna Vacca, Andrea Di~Sorbo, Corrado~A Visaggio, and Gerardo Canfora.
\newblock A systematic literature review of blockchain and smart contract
  development: Techniques, tools, and open challenges.
\newblock {\em Journal of Systems and Software}, 174:110891, 2021.

\bibitem{tortoise}
Yanhao Wang, Xiangkun Jia, Yuwei Liu, Kyle Zeng, Tiffany Bao, Dinghao Wu, and
  Purui Su.
\newblock Not all coverage measurements are equal: Fuzzing by coverage
  accounting for input prioritization.
\newblock In {\em NDSS}, 2020.

\bibitem{gf1}
Cheng Wen, Haijun Wang, Yuekang Li, Shengchao Qin, Yang Liu, Zhiwu Xu, Hongxu
  Chen, Xiaofei Xie, Geguang Pu, and Ting Liu.
\newblock Memlock: Memory usage guided fuzzing.
\newblock In {\em Proceedings of the ACM/IEEE 42nd International Conference on
  Software Engineering}, pages 765--777, 2020.

\bibitem{defigrowth}
Sam Werner, Daniel Perez, Lewis Gudgeon, Ariah Klages-Mundt, Dominik Harz, and
  William Knottenbelt.
\newblock Sok: Decentralized finance (defi).
\newblock In {\em Proceedings of the 4th ACM Conference on Advances in
  Financial Technologies}, pages 30--46, 2022.

\bibitem{harvey}
Valentin W{\"u}stholz and Maria Christakis.
\newblock Harvey: A greybox fuzzer for smart contracts.
\newblock In {\em Proceedings of the 28th ACM Joint Meeting on European
  Software Engineering Conference and Symposium on the Foundations of Software
  Engineering}, pages 1398--1409, 2020.

\bibitem{llm7}
Chunqiu~Steven Xia, Matteo Paltenghi, Jia~Le Tian, Michael Pradel, and Lingming
  Zhang.
\newblock Universal fuzzing via large language models.
\newblock {\em arXiv preprint arXiv:2308.04748}, 2023.

\bibitem{repair1}
Chunqiu~Steven Xia and Lingming Zhang.
\newblock Conversational automated program repair.
\newblock {\em arXiv preprint arXiv:2301.13246}, 2023.

\bibitem{prompt}
JD~Zamfirescu-Pereira, Richmond~Y Wong, Bjoern Hartmann, and Qian Yang.
\newblock Why johnny can’t prompt: how non-ai experts try (and fail) to
  design llm prompts.
\newblock In {\em Proceedings of the 2023 CHI Conference on Human Factors in
  Computing Systems}, pages 1--21, 2023.

\bibitem{llm1}
Cen Zhang, Mingqiang Bai, Yaowen Zheng, Yeting Li, Xiaofei Xie, Yuekang Li, Wei
  Ma, Limin Sun, and Yang Liu.
\newblock Understanding large language model based fuzz driver generation.
\newblock {\em arXiv preprint arXiv:2307.12469}, 2023.

\bibitem{llmsurvey}
Wayne~Xin Zhao, Kun Zhou, Junyi Li, Tianyi Tang, Xiaolei Wang, Yupeng Hou,
  Yingqian Min, Beichen Zhang, Junjie Zhang, Zican Dong, et~al.
\newblock A survey of large language models.
\newblock {\em arXiv preprint arXiv:2303.18223}, 2023.

\bibitem{taint1}
Xiaogang Zhu, Sheng Wen, Seyit Camtepe, and Yang Xiang.
\newblock Fuzzing: a survey for roadmap.
\newblock {\em ACM Computing Surveys (CSUR)}, 54(11s):1--36, 2022.

\end{thebibliography}

\end{document}